# Parameter security characterization of knapsack public-key crypto under quantum computing

Xiangqun Fu[1,2], Wansu Bao[1,2,*], Jianhong Shi[1,2], Fada Li[1,2], Yuchao Zhang[1,2]

([1]*Zhengzhou Information Science and Technology Institute, Zhengzhou, China 450004*
[2]*Synergetic Innovation Center of Quantum Information and Quantum Physics, University of Science and Technology of China, Hefei, China 230026*)

**Abstract**: In order to research the security of the knapsack problem under quantum algorithm attack, we study the quantum algorithm for knapsack problem over $Z_r$ based on the relation between the dimension of the knapsack vector and $r$. First, the oracle function is designed based on the knapsack vector $B = (b_1, b_2, \cdots, b_n)$ and $S$, and the quantum algorithm for the knapsack problem over $Z_r$ is presented. The observation probability of target state is not improved by designing unitary transform, but oracle function. Its complexity is polynomial. And its success probability depends on the relation between $n$ and $r$. From the above discussion, we give the essential condition for the knapsack problem over $Z_r$ against the existing quantum algorithm attacks, i.e. $r < O(2^n)$. Then we analyze the security of the Chor-Rivest public-key crypto.

## I. INTRODUCTION

Knapsack public-key encryption schemes[1] are based on the knapsack problem, which is NP-complete. Merkle-Hellman knapsack encryption scheme was the first concrete realization of a public-key encryption scheme. As its secure basis is superincreasing knapsack problem, it has been demonstrated to be insecure. Many variations have subsequently been proposed, whose knapsack vector density are less than 1. $L^3$-lattice basis reduction algorithm[2] is a polynomial-time algorithm for finding a reduced basis when given a basis for a lattice. In 1991, Schnorr and Euchner presented and improved algorithm[3]. In 1992, Coster et al gave an algorithm for low density knapsack problem based on lattice basis reduction algorithm[4]. If the density of the knapsack is less than 0.9408, the knapsack problem can be solved with high probability. Thus most variations of the Merkle-Hellman scheme are insecure. However, there is not an efficient algorithm for solving knapsack problem, and the complexity of the best algorithm is $O(n 2^{n/2})$ [5], where $n$ is the dimension of the knapsack vector.

Shor presented a quantum algorithm for order-finding [6], based on which factoring and discrete logarithm over finite field can be solved in polynomial time. The public-key crypto, such as RSA and ECC, is under threat. Grover presented a quantum search algorithm achieving quadratic speedup for the unstructured search problem[7], which can be used for searching cryptographic key. From then on, lots of achievements have been made on the quantum computer and quantum algorithm[8~11].

The success probability of the quantum algorithm is depicted by the measurement probability of the target state(MPTS). In the existing quantum algorithm, the MPTS is improved by designing unitary transform. For example, running quantum Fourier transform once can obtain high MPTS in Shor's algorithm; running Grover iteration $O(\sqrt{N})$ times can obtain high MPTS in Grover's algorithm, where $N$ is the size of the search space. Therefore the success probability and complexity of the quantum algorithm depend on the MPTS. Most of the polynomial quantum algorithms can be reduced to Abelian hidden subgroup problem, but knapsack problem cannot. Thus, hoe to design the quantum algorithm for knapsack problem based on new method of improving MPTS requires further study.

The best quantum algorithm for general knapsack problem was presented by V.Arvind and R.Schuler[12], which is based on the 0-1 Integer Linear Programs and Grover's algorithm. Its complexity and success probability are respective $O(c2^{n/3})$ and 1. There is no efficient quantum algorithm for knapsack problem. Thus we conjecture that knapsack problem can resist the quantum computing attack.

Since the knapsack public-key crypto is based on the knapsack problem with special property, such as superincreasing and low density, it can be broken in polynomial. For the classical algorithm for knapsack problem over $Z_r$, its complexity is correlated with $n$ (the dimension of the knapsack vector $B = (b_1, b_2, \cdots, b_n)$), not with $r$. However, the relationship between $r$ and the quantum algorithm for

---

knapsack problem over $Z_r$ is still unknown. Thus, based on the relation between $n$ and $r$, it is worth giving the essential condition for the knapsack problem over $Z_r$ against the existing quantum algorithm attacks. Furthermore, it can provide the theoretical basis for the security of the public-key crypto, which is based on the NPC problem.

In this paper, we present a quantum algorithm for knapsack problem over $Z_r$, based on which the essential condition for the knapsack problem over $Z_r$ against this algorithm is given. Then we analyze the security of Chor-Rivest public-key crypto.

## II. BACKGROUND

The security of the knapsack public-key crypto depends on knapsack problem, which is NPC problem.

**Definition 1** The *knapsack problem* [5] is the following: given a vector $B = (b_1, b_2, \cdots, b_n)$ and $S$, determine whether or not there exist $x_i \in \{0,1\}$, $1 \leq i \leq n$, such that $\sum_{i=1}^{n} b_i x_i = S$.

Knapsack public-key crypto is generally based on knapsack problem over $Z_r$.

**Definition 2** The *knapsack problem over* $Z_r$ is the following: given a vector $B = (b_1, b_2, \cdots, b_n)$ and $S$, where $b_i \in Z_r$, determine whether or not there exist $x_i \in \{0,1\}$, $1 \leq i \leq n$, such that $\sum_{i=1}^{n} b_i x_i \bmod r = S$. $B$ is called the knapsack vector, whose dimension is $n$.

In quantum algorithm, quantum Fourier transform is a very important unitary transform, which can be used for preparing superposition state and improving the MPTS.

**Definition 3** (*Quantum Fourier transform*)[13] The quantum Fourier transform QFT on an orthonormal basis $|0\rangle, |1\rangle, \cdots, |N-1\rangle$ is defined to be a linear operator $U_F$ with an action on the basis states described by

$$U_F : |j\rangle \mapsto \frac{1}{\sqrt{N}} \sum_{k=0}^{N-1} \omega_N^{jk} |k\rangle.$$

Therefore, the action on an arbitrary state may be written as

$$U_F : \sum_{j=0}^{N-1} x_j |j\rangle \mapsto \frac{1}{\sqrt{N}} \sum_{k=0}^{N-1} \sum_{j=0}^{N-1} x_j \omega_N^{jk} |k\rangle,$$

where the symbol "$\mapsto$" stands for an invertible transform and $\omega_N = e^{2\pi i/N}$. If $N = 2^n$, it is called the $n$-dimensional QFT.

## III. QUANTUM ALGORITHM FOR KNAPSACK PROBLEM OVER $Z_r$

For the classical algorithm for knapsack problem over $Z_r$, its complexity is correlated with $n$ (the dimension of the knapsack vector $B = (b_1, b_2, \cdots, b_n)$), not with $r$. According to Definition 1 and Definition 2, the bigger $r$ is, i.e. the less the number of solutions for knapsack problem over $Z_r$ is, the higher safety of the knapsack public-key crypto is. Thus, we present a quantum algorithm for knapsack problem over $Z_r$, which can be used for depicting the security of knapsack problem over $Z_r$ based on the relation between $r$ and $n$.

For arbitrary $x$ and $y$, $z = x \oplus y$ means $z_i = (x_i + y_i) \bmod 2$, where $(x_1, x_2, \cdots, x_n)$, $(y_1, y_2, \cdots, y_n)$, $(z_1, z_2, \cdots, z_n)$ are respectively the binary representation of $x$, $y$ and $z$, i.e. $x = \sum_{i=1}^{n} 2^{n-i} x_i$, $y = \sum_{i=1}^{n} 2^{n-i} y_i$ and $z = \sum_{i=1}^{n} 2^{n-i} z_i$.

For the sake of argument, $k$ is denoted as the number of the solutions for knapsack problem, i.e. there are $k$ times $(m_{i1}, m_{i2}, \cdots, m_{in}) \in Z_2^n$, such that $\sum_{j=1}^{n} m_{ij} b_i = S$, where $i = 1, 2, \cdots, k$ and $(m_{i1}, m_{i2}, \cdots, m_{in}) \in Z_2^n$ is the binary representation of $m_i$.

**Theorem 1** Suppose $f(x, y) = 2\sum_{i=1}^{n}(x_i \oplus y_i)b_i + S - \sum_{i=1}^{n} x_i b_i$ and $g(z) = \sum_{i=1}^{n} 2z_i b_i$, where $x, y, z \in \{0, 1, \cdots, 2^n\}$, $x = \sum_{i=1}^{n} 2^{n-i} x_i$, $y = \sum_{i=1}^{n} 2^{n-i} y_i$ and $z = \sum_{i=1}^{n} 2^{n-i} z_i$, then $f(x, y) = g(z)$ with $z_i = x_i \oplus y_i$ if and only if

$$x_i = m_{ji}, \quad y_i = z_i \oplus m_{ji}.$$

**Proof:** If $f(x, y) = g(z)$,

$$2\sum_{i=1}^{n}(x_i \oplus y_i)b_i + S - \sum_{i=1}^{n} x_i b_i = \sum_{i=1}^{n} 2z_i b_i$$

i.e.

$$S = \sum_{i=1}^{n}(2z_i + x_i - 2(x_i \oplus y_i))b_i$$

Since $z_i = x_i \oplus y_i$, $2z_i + x_i - 2(x_i \oplus y_i)$ is either 0 or 1.

Furthermore, $j$ can be found, such that

$$2z_i + x_i - 2(x_i \oplus y_i) = m_{ji}$$

Case 1. When $z_i = 0$, $x_i = 2(x_i \oplus y_i) + m_{ji}$.

Case 1(a). If $m_{ji} = 0$, $\{x_i = 0, y_i = 0\}$, i.e. $x_i = m_{ji}$ and $y_i = z_i \oplus m_{ji}$.

Case 1(b). If $m_{ji} = 1$, $\{x_i = 1, y_i = 1\}$, i.e. $x_i = m_{ji}$ and $y_i = z_i \oplus m_{ji}$.

Case 2. When $z_i = 1$, $x_i + 2 = 2(x_i \oplus y_i) + m_{ji}$.

Case 2(a). If $m_{ji} = 0$, $\{x_i = 0, y_i = 1\}$, i.e. $x_i = m_{ji}$ and $y_i = z_i \oplus m_{ji}$.

Case 2(b). If $m_{ji} = 1$, $\{x_i = 1, y_i = 0\}$, i.e. $x_i = m_{ji}$ and $y_i = z_i \oplus m_{ji}$.

If there is $j$, such that

$$x_i = m_{ji} \text{ and } y_i = z_i \oplus m_{ji}$$

it is obvious that $f(x, y) = g(z)$.

Thus we obtain Theorem 1.

According to Theorem 1, we can obtain Corollary 1.

**Corollary 1** Suppose $f(x, y) = 2\sum_{i=1}^{n}(x_i \oplus y_i)b_i + S - \sum_{i=1}^{n} x_i b_i$ and $h(u, v) = \sum_{i=1}^{n} 2(u_i \oplus v_i)b_i$, where $x, y, u, v \in \{0, 1, \cdots, 2^n\}$, $x = \sum_{i=1}^{n} 2^{n-i} x_i$, $y = \sum_{i=1}^{n} 2^{n-i} y_i$, $u = \sum_{i=1}^{n} 2^{n-i} u_i$ and $v = \sum_{i=1}^{n} 2^{n-i} v_i$, then $f(x, y) = h(u, v)$ with $u_i \oplus v_i = x_i \oplus y_i$ if and only if

$$x_i = m_{ji}, \quad y_i = u_i \oplus v_i \oplus m_{ji}.$$

$f$ and $h$ is called the adjoint function. According to Corollary 1, if $x, y, u, v$ satisfy $u_i \oplus v_i = x_i \oplus y_i$ and $f(x, y) = h(u, v)$, $(x_1, x_2, \cdots, x_n)$ is a solution of knapsack problem over $Z_r$.

**Definition 4** The *extended knapsack problem* over $Z_r$ is the following: given a vector $B = (b_1, b_2, \cdots, b_n)$ and $S'$, where $b_i \in Z_r$, determine whether or not there exist $x_i' \in \{-1, 0, 1, 2\}$, $1 \leq i \leq n$, such that $\sum_{i=1}^{n} b_i x_i' \mod r = S'$.

Let $k'$ be the number of solutions of the extended knapsack problem over $Z_r$, i.e. the number of $(m_{i1}', m_{i2}', \cdots, m_{in}') \in \{-1, 0, 1, 2\}^n$ is $k'$, such that $\sum_{j=1}^{n} m_{ij}' b_j = S'$ for $i = 1, 2, \cdots, k'$.

If $B = \{(m_{i1}', m_{i2}', \cdots, m_{in}') \in \{-1, 0, 1, 2\}^n | i = 1, 2, \cdots, k'\}$ and $B' = \{0, 1, \cdots, r-1\}$, it is obvious that $|B| = 4^n$ and $|B'| = r$. Thus, for arbitrary $S' \in B$, the average number of solutions of the extended knapsack problem over $Z_r$ is

$$4^n / r.$$

Furthermore, if $r = O(4^n)$, i.e. $4^n / r = O(1)$, we can obtain that $k' = O(1)$.

Now, we give the quantum algorithm for the knapsack problem over $Z_r$, where $r$ and $n$ satisfy $r = O(4^n)$. This algorithm can be showed as follows.

**Quantum algorithm for the knapsack problem over $Z_r$**

*Step*1 Give 5 quantum registers, whose dimension are respectively 1, $n$, $n$, $n$ and $n + \lceil \log n \rceil + 3$. Their initial state are all $|0\rangle$. Apply the quantum Fourier transform to the former four registers, then obtain the superposition state

$$|0\rangle|0\rangle|0\rangle|0\rangle|0\rangle$$
$$\mapsto \frac{1}{\sqrt{2^{3n+1}}} \sum_{a=0}^{1} \sum_{x=0}^{2^n-1} \sum_{y=0}^{2^n-1} \sum_{z=0}^{2^n-1} |a\rangle|x\rangle|y\rangle|z\rangle|0\rangle.$$

*Step*2 Perform the oracle function $G(a, x, y, z) = \begin{cases} f(x, y), a = 0 \\ h(x, z), a = 1 \end{cases}$,

$$\frac{1}{\sqrt{2^{3n+1}}} \sum_{a=0}^{1} \sum_{x=0}^{2^n-1} \sum_{y=0}^{2^n-1} \sum_{z=0}^{2^n-1} |a\rangle|x\rangle|y\rangle|z\rangle|0\rangle$$
$$\mapsto \frac{1}{\sqrt{2^{3n+1}}} \sum_{a=0}^{1} \sum_{x=0}^{2^n-1} \sum_{y=0}^{2^n-1} \sum_{z=0}^{2^n-1} |a\rangle|x\rangle|y\rangle|z\rangle|G(a, x, y, z)\rangle$$

where the definition of $f$ and $h$ are the same as Definition 1, $y = \sum_{i=1}^{n} 2^{n-i} y_i$ and $z = \sum_{i=1}^{n} 2^{n-i} z_i$.

*Step*3 Measure the last register and get $A$. We can obtain

$$\frac{1}{\sqrt{2^n (t+t')}} (\sum_{z=0}^{2^n-1} \sum_{x'',y''} |0\rangle|x''\rangle|y''\rangle|z\rangle|A\rangle + \sum_{y=0}^{2^n-1} \sum_{x',z'} |1\rangle|x'\rangle|y\rangle|z'\rangle|A\rangle)$$

where $h(x', z') = A$, $f(x'', y'') = A$, $t = |\{(x', z') | h(x', z') = A\}|$, $t' = |\{(x'', y'') | f(x'', y'') = A\}|$ and $\sum_{i=1}^{n} (2(x_i' \oplus z_i') + x_i'' - 2(x_i'' \oplus y_i'')) b_i = S$.

*Step*4 Measure the first register and get $t^0$. Obtain $\frac{1}{\sqrt{2^n t'}} \sum_{z=0}^{2^n-1} \sum_{x'',y''} |0\rangle|x''\rangle|y''\rangle|z\rangle|A\rangle$ if $t^0 = 0$, otherwise output failure.

*Step*5 Measure the second and third register. We can obtain $x^0$ and $y^0$. Set $m_i^0 = x_i^0$.

*Step*6 $(m_1^0, m_2^0, \cdots, m_n^0)$ is one solution of the knapsack problem over $Z_r$ if $\sum_{i=1}^{n} m_i^0 b_i = S$, otherwise output failure.

Correctness, success probability and complexity are very important in a quantum algorithm.

**Correctness and success probability**

The success probability is unrelated to *Step*1 and *Step*2. Only when $t^0 = 0$ in *Step*4 will the algorithm go on performing *Step*5. It is obvious that *Step*1 and *Step*2 are both correct. Thus the algorithm's correctness and success probability are only related to *Step*3, *Step*4, *Step*5 and *Step*6.

(1) In *Step*3, we can obtain the measurement result $A$.

Case 1. If there are $x'$ and $z'$, such that $A = G(1, x', y, z')$, $|1\rangle|x'\rangle|y\rangle|z'\rangle|A\rangle$ is firmly included in the superposition state and $h(x', z') = A$. According to Corollary 1, $|0\rangle|x''\rangle|y''\rangle|z\rangle|A\rangle$ is also included in the superposition state and $\sum_{i=1}^{n}(2(x_i' \oplus z_i') + x_i'' - 2(x_i'' \oplus y_i''))b_i = S$.

Case 2. If there are $x''$ and $y''$, such that $A = G(0, x'', y'', z)$, $|0\rangle|x''\rangle|y''\rangle|z\rangle|A\rangle$ is firmly included in the superposition state and $f(x'', y'') = A$. According to Corollary 1, $|1\rangle|x'\rangle|y\rangle|z'\rangle|A\rangle$ is also included in the superposition state and $\sum_{i=1}^{n}(2(x_i' \oplus z_i') + x_i'' - 2(x_i'' \oplus y_i''))b_i = S$.

Thus, according to Case 1 and Case 2, we can obtain
$$\frac{1}{\sqrt{2^n(t+t')}}(\sum_{z=0}^{2^n-1}\sum_{x'',y''}|0\rangle|x''\rangle|y''\rangle|z\rangle|A\rangle + \sum_{y=0}^{2^n-1}\sum_{x',z'}|1\rangle|x'\rangle|y\rangle|z'\rangle|A\rangle)$$
in *Step*3.

(2) Since $\sum_{i=1}^{n}(2(x_i' \oplus z_i') + x_i'' - 2(x_i'' \oplus y_i''))b_i = S$, for arbitrary $m_j$,
$$A - S = \sum_{i=1}^{n} 2(x_i' \oplus z_i')b_i - S$$
$$= \sum_{i=1}^{n}(2(x_i' \oplus z_i') - m_{ji})b_i$$
$$= \sum_{i=1}^{n}(2(x_i'' \oplus y_i'') - x_i'')b_i$$

It is obvious that $2(x_i'' \oplus y_i'') - x_i'' \in \{-1, 0, 1, 2\}$ and $2(x_i' \oplus z_i') - m_{ji} \in \{-1, 0, 1, 2\}$. If the number of $D_j = (d_1, d_2, \cdots, d_n)$ is $t''$, where $\sum_{i=1}^{n} b_i d_i \bmod r = A - S$ and $d_i \in \{-1, 0, 1, 2\}$, we will get $t \le t''$ and $t' \le t''$. Since $x''$ and $y''$ are independent of one another, $(2(x_1'' \oplus y_1'') - x_1'', 2(x_2'' \oplus y_2'') - x_2'', \cdots, 2(x_n'' \oplus y_n'') - x_n'')$ can go through $D_1, D_2, \cdots, D_{t''}$. Furthermore, in *Step*4, the probability of $t^0 = 0$ is
$$P_1 = \frac{t'}{t+t'} > \frac{1}{2}$$

(3) In *Step*4, each quantum state $|0\rangle|x''\rangle|y''\rangle|z\rangle|A\rangle$ of the superposition state satisfies
$$A - S = \sum_{i=1}^{n}(2(x_i'' \oplus y_i'') - x_i'')b_i$$

For arbitrary $m_j$, there is $y^j$, such that

$$2(m_{ji} \oplus y_i^j) - m_{ji} = 2(x_i' \oplus z_i') - m_{ji}$$

i.e.

$$\sum_{i=1}^{n}(2(m_{ji} \oplus y_i^j) - m_{ji})b_i = A - S$$

Thus, the superposition state must include quantum state $|0\rangle|m_j\rangle|y^j\rangle|z\rangle|A\rangle$. After measuring in *Step*5, the probability of getting $m_j$ is

$$P_2 = \frac{k}{t'}$$

According to (1), (2) and (3), the success probability of this algorithm is
$$P = P_1 \cdot P_2$$
$$> \frac{1}{2} \cdot \frac{k}{t'}$$
$$\geq \frac{1}{2t'}$$

When $r = O(4^n)$, the average value of $t'$ is $O(1)$, i.e. the probability of this algorithm is at least $\frac{1}{2O(1)}$.

**Complexity**

In *Step*1, the algorithm need take 1 time one-dimensional quantum Fourier transform and 3 times $n$-dimensional quantum Fourier transform. And $n$-dimensional quantum Fourier transform requires $O(n^2)$ elementary quantum gates[13]. Thus, it requires $O(n^2)$ elementary quantum gates for *Step*1.

In *Step*2, the algorithm requires $2n+1$ addition operations, where the addend's length is $n + \lceil \log n \rceil + 1$. And the addition operations requires $O((n + \lceil \log n \rceil + 1)^2)$ elementary quantum gates[14]. Thus, it requires $O((n + \lceil \log n \rceil + 1)^2)$ elementary quantum gates for *Step*2.

In *Step*3 and *Step*4, the algorithm only need take measuring operation. Thus, it requires $O(1)$ elementary quantum gates for *Step*3 and *Step*4.

In *Step*5, the algorithm need take 2 times measuring operation and 1 time classical computing. Thus, it requires $O(1)$ elementary quantum gates for *Step*5. The classical computing can be done in polynomial time.

In *Step*6, the algorithm need take 1 time classical computing, which can be done in polynomial time.

Thus, it requires $O((n + \lceil \log n \rceil + 1)^2)$ for this algorithm, whose complexity is polynomial.

If adjoint function $f(x, y)$ and $h(u, v)$ are equal, the relation between $m_j$, $x$ and $y$ can be got. Based on the relation, we can construct a system of $n$ linear equations in $n$ variables $m_{j1}, m_{j2}, \cdots, m_{jn}$, whose rank is $n$. Further, we can obtain $m_j$.

**Safe parameter of knapsack problem under quantum computing**

The success probability of the quantum algorithm for knapsack problem over $Z_r$ depends on the number of solutions of the extended knapsack problem over $Z_r$. If $t' = O(2^n)$, i.e. the success probability of the algorithm is $1/O(2^n)$, the algorithm need implementing $O(2^n)$ times to obtain one solution of the knapsack problem over $Z_r$ with $O(1)$ probability. Thus, to guarantee the security of the knapsack public-key crypto, $r$ and $n$ must satisfy $r < O(2^n)$, i.e. $4^n/r > O(2^n)$. Furthermore, the bigger $r$ is, the higher the success probability of the algorithm, i.e. the easier the knapsack problem over $Z_r$ can be solved.

NPC problem is the secure basis of the post-quantum public-key crypto. Since all NPC problems are reduced each other, proper parameters are chosen for the post-quantum public-key crypto to resist the quantum algorithm for knapsack problem over $Z_r$, which based on the NPC problem.

## IV. THE SECURITY OF CHOR-RIVEST KNAPSACK PUBLIC-KEY CRYPTO

Merkle-Hellman knapsack encryption scheme is insecure, most of whose variations are also insecure. Since knapsack vector density of these variations are less than 1, they cannot resist the $L^3$ algorithm. Chor-Rivest knapsack public-key crypto is the only known public-key crypto scheme that does not use some form of modular multiplication[15]. And it can resist the $L^3$ algorithm, since its knapsack vector density is more than 1. It is shown as follows.

**Chor-Rivest knapsack public-key crypto**[15]

*Key generation*

*Step*1 Choose finite field $F_q$ of characteristic $p$, where $q = p^h$ and $p \geq h$, and for which the discrete logarithm problem is feasible.

*Step*2 Choose a random monic irreducible polynomial $f(x)$ of degree $h$ over $Z_p$. The elements of $F_q$ will be represented as polynomials in $Z_p[x]$ of degree less than $h$, with multiplication performed modulo $f(x)$.

*Step*3 Select a random primitive element $g(x)$ of $F_q$.

*Step*4 Compute $a_i = \log_{g(x)}(x+i)$, for each $i \in Z_p$.

*Step*5 Choose a random permutation $\pi$ on $(0,1,\cdots,p-1)$.

*Step*6 Choose $d$, $0 \leq d \leq p^h - 2$.

*Step*7 Compute $b_i = (a_{\pi(i)} + d) \mod (p^h - 1), 0 \leq i \leq p-1$.

*Step*8 Public key is $((b_0, b_1, \cdots, b_{p-1}), p, h)$; private key is $(f(x), g(x), \pi, d)$.

*Encryption*. B should do the following.

*Step*1 Transform the message $m$ into a binary vector $M = (M_0, M_1, \cdots, M_{p-1})$ of length $p$ having exactly $h$ 1's as follows:
  i. Let $l \leftarrow h$.
  ii. For $i$ from 1 to $p$ do the following:
     If $m \geq \binom{p-i}{l}$ then set $M_{i-1} \leftarrow 1$, $m \leftarrow m - \binom{p-i}{l}$, $l \leftarrow l-1$. Otherwise, set $M_{i-1} \leftarrow 0$.

*Step*2 Compute $c = \sum_{i=0}^{p-1} M_i b_i \mod (p^h - 1)$.

*Step*3 Send the ciphertext $c$ to.

*Decryption*. A should do the following.

*Step*1 Compute $r = (c - hd) \mod (p^h - 1)$.

*Step*2 Compute $u(x) = g(x)^r \mod f(x)$.

*Step*3 Compute $s(x) = u(x) + f(x)$.

*Step*4 Factor $s(x)$ into $s(x) = \prod_{j=1}^{h}(x + t_j)$, where $t_j \in Z_p$.

*Step*5 Compute a binary vector $M = (M_0, M_1, \cdots, M_{p-1})$ as follows. The components of $M$ that are 1 have indices $\pi^{-1}(t_j)$. The remaining components are 0.

*Step*6 The message $m$ is recovered from $M$ as follows:
  i. Let $m \leftarrow 0$, $l \leftarrow h$.
  ii. For $i$ from 1 to $p$ do the follows:

If $M_{i-1} = 1$ then set $m \leftarrow m + \binom{p-i}{l}$ and $l \leftarrow l - 1$.

The correctness of Chor-Rivest scheme is shown in Reference [15].

If we obtain $M$ from $c, b_0, b_1, \cdots, b_{p-1}$, it's easy to obtain $m$. The security of Chor-Rivest scheme is based on the knapsack problem over $Z_{p^h-1}$.

To date, under classical computing, safe parameters $p$ and $h$ for the Chor-Rivest scheme were presented in Reference [16], which should satisfy five conditions, i.e.

(1) $p$ is a prime,
(2) $h$ is a prime,
(3) $h \leq p$,
(4) $11 \leq h \leq 31$,
(5) $10^{44} < p^h - 1 < 10^{60}$.

In this paper, these conditions are called Five Conditions (FC) for short. Furthermore, the greatest prime factor of $p^h - 1$ is equal to or less than $10^{13}$ for computational feasibility.

If $p = 109$ and $h = 29$, which are the proposed parameters in Reference [16] satisfy the FC. The Chor-Rivest scheme for these parameters is secure under classical computing.

Since $4^{109}/(109^{29} - 1) \approx 3460753.1 < O(2^{109})$, the Chor-Rivest scheme for these parameters can be broken with at least $1/6921506.2$ probability by the quantum algorithm for knapsack over $Z_r$, i.e. it can be broken with ignorable probability in polynomial time.

Thus, under quantum computing, FC and $4^p/(p^h - 1) > O(2^p)$ can be used for depicting the security of Chor-Rivest scheme.

## V. CONCLUSION

The quantum algorithm for knapsack problem over $Z_r$ is presented, whose success probability relies on the number of solutions of the extended knapsack problem over $Z_r$. The method of parameter selection is given for the knapsack problem over $Z_r$ against the existing quantum computing. However, the quantum algorithm only solves the knapsack problem over $Z_r$ with special parameter. How to design a quantum algorithm for general knapsack problem over $Z_r$ requires further study.

This work was supported by the National Basic Research Program of China (Grant No. 2013CB338002).